# $^3$He in stars of low and intermediate mass

A. Weiss[1], J. Wagenhuber[1], and P.A. Denissenkov[2,1]

[1] Max-Planck-Institut für Astrophysik, Karl-Schwarzschild-Str. 1, 85740 Garching, Federal Republic of Germany
[2] Astronomical Institute of the St. Petersburg University, Bibliotechnaja Pl. 2, Petrodvorets, 198904 St. Petersburg, Russia

**Abstract.** We follow the evolution of $^3$He in stars of mass $0.8 \cdots 10 M_\odot$ for Pop. I and II compositions from the main sequence until advanced stages on the AGB. Under standard assumptions we confirm earlier results of more restricted investigations that low-mass stars up to $5 M_\odot$ are net producers of $^3$He. We show that the inclusion of additional mixing due to diffusion beneath the convective envelope simultaneously leads to observed carbon isotope anomalies observed in globular cluster Red Giants and to a strong reduction of $^3$He, such that stars exhibiting such anomalies will have destroyed $^3$He contrary to the standard picture.

Therefore, knowledge about these abundances enables us to determine the baryon density $\Omega_B$, which immediately leads to results about kinds and amounts of dark matter and other cosmological questions. A direct measurement of almost primordial abundances may be possible, but usually is very difficult and results are controversial. Examples are the determination of $^4$He in low-metallicity galaxies (see, e.g., Izotov, Trinh & Lipovetsky 1994 and Olive & Steigman 1995) and of D in the $z = 3.32$ quasar Q0014+813 (Carswell et al. 1994; Songaila et al. 1994). Solar system or galactic determinations are much easier and reliable, but they suffer from the fact that galacto-chemical processes have altered the light element abundances. To extrapolate back to primordial values therefore requires knowledge and models about those processes.

Most interesting and complex in this respect is $^3$He for several reasons. Firstly, because D is converted completely to $^3$He in stars via proton capture at temperatures of about $6\,10^5$ K, which are already reached in completely convective and homogeneous pre-main sequence stars. Extrapolating back to the primordial D abundance therefore requires measurements of D as well as $^3$He plus knowledge about the fate of $^3$He. A simplified relation used is (Yang et al. 1984)

$$\left(\frac{D}{H}\right)_p \leq \left(\frac{D + {}^3\mathrm{He}}{H}\right)_p \leq \left(\frac{D}{H}\right)_t + \frac{1}{g_3}\left(\frac{{}^3\mathrm{He}}{H}\right)_t,$$

where the subscript $p$ refers to primordial and $t$ to abundances at a later time, e.g. pre-solar. $g_3$ is the fraction of $^3$He that survived processing in stars and was returned to the ISM. Steigman (1994) has used this relation to conclude that solar $^3$He and D abundances already prohibit the high Q0014+813 D values, when $g_3 \geq 0.25$ (Dearborn, Schramm & Steigman 1986) is assumed. This is even a very conservative estimate, since the value for $g_3$ was actually derived from massive stars alone; Dearborn et al. (1986) estimate $g_3 \geq 0.5$ for stars with $M \geq 0.8 M_\odot$ under the very conservative assumption that $g_3 \approx 1$ for stars below $3 M_\odot$. The higher $g_3$, the lower the allowed $D_p$.

Besides the fact that $^3$He is the result of stellar D processing, its fate in stars is interesting in itself. Not only is all primordial D converted to $^3$He, but $^3$He is also produced in considerable quantities during hydrogen burning via the p-p chain: two protons react to D, which is immediately burned to $^3$He via a further proton capture. In the complete p-p chain the destructive reactions are

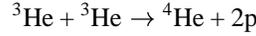
$$^3\mathrm{He} + {}^3\mathrm{He} \to {}^4\mathrm{He} + 2p$$

and

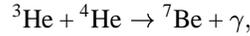
$$^3\mathrm{He} + {}^4\mathrm{He} \to {}^7\mathrm{Be} + \gamma,$$

which, however, proceed very fast only at temperatures above $\approx 1.5 \cdot 10^7$K. Consequently, stars have regions where $^3$He is produced and others, where it is effectively destroyed, because the equilibrium abundance of $^3$He in stellar layers burning hydrogen via the p-p cycle is much lower than the primordial one. The task for stellar evolution theory is therefore to predict how much $^3$He is produced, can survive the subsequent evolution and be returned to the ISM. Presently, stars more massive than $\approx 5 M_\odot$ are considered to be net destroyers of $^3$He (Dearborn et al. 1986), while less massive ones are net producers. Under standard assumptions about the star formation rate (SFR), the initial mass function (IMF) and further details of galactic evolution a severe overproduction of $^3$He with respect to the measured pre-solar and ISM $^3$He abundance is obtained (Galli et al. 1995, see this paper also for a nice compilation of recent measurements of D and $^3$He; Olive et al. 1995). This is mainly due to the strong net production of $^3$He in low-mass stars, as already recognized by Rood, Steigman & Tinsley (1976). However, there are very few papers in the literature, which investigated the problem of $^3$He production in low-mass stars specifically; most values quoted are derived from the early results of Iben (e.g. Iben 1966, 1967), who followed the stars only up to the first thermal pulse and used a reaction rate for $^3$He($^3$He, 2p)$^4$He too small by a factor 5. Later on, calculations by Rood (1972), Rood et al. (1976) and Sackmann, Smith & Despain (1974) were mainly concerned with effects on the first Red Giant branch. Galli et al. (1995) also use some data from Vassiliadis & Wood (1993) for Pop. II models that experienced a number of thermal pulses and from Straniero (unpublished) without any details about the computations being available. The results of such calculations about the abundance of $^3$He are used as input for the galacto-chemical models (Truran & Cameron 1971; Galli et al. 1995). Recently, Galli et al. (1994) have proposed to reduce $^3$He-production in low-mass stars by postulating a low-energy resonance in the $^3$He-$^3$He reaction.

The present paper is a systematic investigation into this problem, intended to yield detailed and specific results about $^3$He in low- and intermediate-mass stars. We have put particular emphasis on the AGB phase, which previously has not been investigated in this context except for the more global work by Vassiliadis & Wood (1993). In sect. 2 the computational method will shortly be described; then the results will follow in sect. 3. There, we will discuss in detail the various phases and regions of $^3$He production and destruction in stars of 5 and 1.25 $M_\odot$. Because the yield of $^3$He depends very strongly on the mass loss history, this is important for future use of our results with different assumptions about mass loss or envelope ejection. In sect. 4, we will add results from non-standard calculations that result in a net destruction of $^3$He, before our summary follows.

All stellar evolution calculations presented in the next section were done with our MPA evolutionary code in the latest version that has been described in Wagenhuber & Weiss (1994a). This version has been designed specifically to allow the accurate, reliable, and stable calculation of thermal pulses on the Asymptotic Giant Branch (AGB). It is therefore most suited to clarify the influence of AGB-evolution on the abundance on $^3$He. For example, stars more massive than $5M_\odot$ experience the so-called hot bottom burning (HBB) of the convective envelope, which could result in the depletion of $^3$He.

All calculations are started on the Zero Age Main Sequence. The composition of the Pop. I models is assumed to be $(X, Y, Z) = (0.70, 0.28, 0.02)$ with solar abundance ratios for the metals. In particular, the initial $^3$He abundance is given by $X_{3,p} = 1.4\,10^{-4}\,Y + 3.0\,10^{-5}\,X$ (Anders & Grevesse 1989), where the second term corresponds to the pre-solar D assumed to be converted completely to $^3$He already on the pre-main sequence (Galli et al. 1995). The initial $^3$He abundance for Pop. I stars is therefore $6.02\,10^{-5}$ in our calculations (for comparison, Galli et al. 1995 derive $6.7\,10^{-5}$). For Pop. II stars, we take $(X, Y, Z) = (0.7499, 0.25, 0.0001)$ and the same relation for the initial $^3$He abundance, which results in $X_{3,p} = 5.75\,10^{-5}$. The latter assumption might not be justified; one could take primordial or solar-scaled abundances instead. We will discuss the influence of the initial abundances on the final $^3$He yields in the next section.

The evolution of $^3$He and other hydrogen burning isotopes is followed by a nuclear reaction network, in which only $\beta$-decays are assumed to be in equilibrium. Abundance changes due to nuclear burning are integrated via a backward differencing scheme with self-adjusting timesteps, which usually are smaller by a factor of ten as compared to the timesteps between two evolutionary models. The reaction rates are taken from Caughlan & Fowler (1988). The uncertainty in the rates concerning $^3$He is always less than 10% and is not influencing our conclusions.

We have followed the evolution of stars of 1, 1.5, 2, 3, 5, 7 and 10 $M_\odot$ for a Pop. I and of 0.8, 1, 1.25, 1.5, 2, 3, 4, 5 $M_\odot$ for a Pop. II composition until several thermal pulses had been experienced such that the effect of AGB evolution could be seen. (The 10 $M_\odot$ model was evolved until the end of core helium burning.) Our AGB models do not experience significant third dredge-up. Obviously, the existence of carbon stars, even for low AGB luminosities, indicate additional mixing which is not included in our standard physics approach. Such questions will be discussed in sect. 4.

The only mixing process we are considering in these calculations is convective mixing, which is assumed to be instantaneous. Convective nuclear-burning regions are processed and mixed in each nuclear timestep. Convection is treated by standard mixing length theory with a ratio of mixing length to scale height of 1.5, which is close to the value needed for a solar model calculated by this version of our code. Effects like semiconvection or overshooting are ignored. We use OPAL opacities (Rogers & Iglesias 1992) supplemented by Los Alamos opacities for low temperatures (Weiss, Keady & Magee 1990). Our equation of state is of a Saha-type with degeneracy taken properly into account. Tests have shown that it agrees very well with the MHD-EOS (Mihalas, Däppen & Hummer 1988) for solar-like conditions.

Of particular interest is the mass loss. Since ab initio mass loss theories are still lacking in stellar evolution theory, and in particular for cool stars, one has to resort to parametrized empirical mass loss formulae. This introduces a certain quality of arbitrariness into the models and into the derived amounts of $^3$He escaping from the stellar processing sites. The same is true for the loss of stellar envelopes due to stellar explosions (for massive stars) or planetary nebulae ejection (for intermediate mass stars), although rather tight bounds for the remaining cores exist, which are inert in the sense of chemical evolution. In our calculations we have used a Reimers law for mass loss: $\dot{M} = -\eta RL/M$ with $\eta = 0.25$. This choice results in an almost negligible mass loss until the tip of the RGB and a very small one of order 1% of the initial mass during the AGB phase. For the ejected envelopes at the end of the evolution we simply assume that the complete hydrogen layers above the hydrogen burning shell are returned to the ISM. Later we will see how important these assumptions are. We will present detailed data about $^3$He enrichment of the stellar envelopes in various stages of evolution, that will allow the calculation of enrichment factors for the ISM under different assumptions about mass loss.

## 2. Results

### 2.1. 5 $M_\odot$, Pop. I model

Ignoring the $^3$He $+^4$He reaction for the moment, the nuclear reaction equation for $^3$He in stellar interiors reads

$$\frac{d^3\text{He}}{dt} = \lambda_{pp}\frac{\text{H}^2}{2} - 2\lambda_{33}\frac{(^3\text{He})^2}{2}$$

where the chemical symbols denote particle densities and the $\lambda$s are the reaction rates per particle pairs (cf. Clayton 1968, pp. 283). As long as temperatures are high enough ($T > 10^7$ K) the equilibrium abundance of $^3$He can be reached within main sequence lifetimes. The $^3$He equilibrium abundance increases with decreasing temperature such that main sequence stars develop a characteristic $^3$He profile as the one shown in Fig. 1 (labelled "MS") for a 5 $M_\odot$ Pop. I model. At the center ($T = 2.8\,10^7$ K) the $^3$He mass fraction is as low as $10^{-7}$ and rises to $8\,10^{-5}$ at $M_r = 3.4M_\odot$ ($T = 1.2\,10^7$ K). Note that the full equilibrium because the pp-cycle is not completed (the burning core extends out to $M_r \approx 2.5 M_\odot$, the convective core to $M_r = 1.2 M_\odot$). Since the time required to reach the $^3$He equilibrium value at a given temperature becomes longer than the main sequence lifetime below a certain temperature, the $^3$He profile develops this characteristic maximum. In the central regions the $^3$He abundance is at its equilibrium value, which, for the larger part is below the initial value. Outside the point of $T \approx 10^7$ K $^3$He is enhanced because of the fact that the equilibrium abundance at those temperatures is always larger than $\approx 10^{-4}$ and therefore higher than the initial value. The integral over the $^3$He profile in Fig. 1 yields a mean stellar $^3$He mass fraction of $3.73\,10^{-5}$, i.e. the star as a whole has destroyed $^3$He. How much the $^3$He abundance will finally have increased depends on the main sequence lifetime of the star; this favours $^3$He production in low-mass stars. The $^3$He maximum is growing further with elapsing time until the end of the main sequence phase (see "TO" in Fig. 1). During the 1$^{\text{st}}$ dredge-up this additional $^3$He will be mixed into the convective envelope. The total amount of $^3$He produced in stars depends on the gradient $dT/dM_r$, which for main sequence stars is very flat and therefore rather suited for $^3$He production. In fact, *basically all $^3$He produced in stars is due to nuclear fusion during the main sequence phase.*

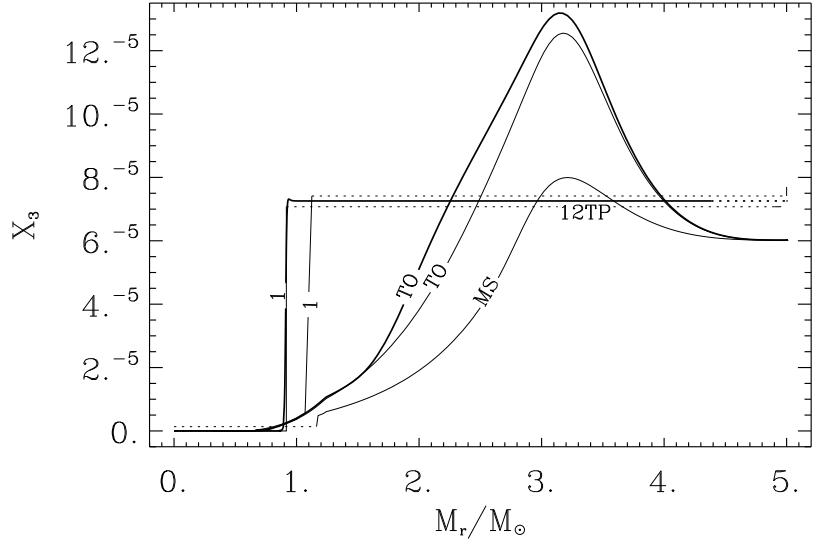

**Fig. 1.** $^3$He profile of a 5 $M_\odot$ Pop. I star in various stages of its evolution until the 12th thermal pulse. $X_3$ denotes the $^3$He mass fraction. Labels along the curves denote models on the early main sequence ("MS"), before and after the turn-off (TO), the first (1) dredge-up and after the last thermal pulse (12TP), when the calculations were stopped. Dotted parts mark convective regions.

The further development of the temperature profile is shown in Fig. 2. For the largest part, i.e. for all regions outside the convective hydrogen core of the main sequence model (thin solid line "MS"; dotted: convective core), temperature is decreasing with time. Already at the end of the main sequence phase (thin solid line "TO") temperatures are dropping and after the turn-off (thick solid line "TO") have fallen below the critical temperature of $10^7$K such that no further $^3$He production can be expected. (Additionally, the subsequent evolution will proceed on a shorter timescale.) On the other hand, parts of the former core become hotter (see, e.g. the lines labelled "1" and "2", which refer to models before and after the 1$^{\text{st}}$ and 2$^{\text{nd}}$ dredge-up). However, the mass fraction extending over temperatures favouring the build-up of high $^3$He equilibrium abundances is very small ($< 10^{-2}\,M_\odot$) such that there is only a minor amount of synthesized $^3$He; furthermore, temperature continues to rise (see the evolution between 1$^{\text{st}}$ and 2$^{\text{nd}}$ dredge-up at $M_r \lesssim 4 M_\odot$) and $^3$He will follow to lower equilibrium values.

Most of the $^3$He which was produced during the main sequence phase and "frozen in" afterwards can reach the photosphere when the star develops a deep convective envelope (see "1" in Fig. 1, showing the downward extending convective regions as dotted parts and in Fig. 2 for the resulting changes in the $^3$He abundance). From now on, the question of how much $^3$He is returned to the ISM depends on the competition between mass loss and increasing temperature at the bottom of the convective envelope (see the curves "2" and "12TP"), which tends to burn $^3$He and thereby reduce its abundance in the whole envelope. A complicating addition to this lies in the occurance of thermal pulses. We found that the maximum envelope $^3$He abundance is reached during the first dredge-up and is $X_{3,\text{e}} = 8.83\,10^{-5}$ (rel. mass fraction). Matter lost at that stage would therefore be enriched by 47%. At the tip of the Red Giant Branch (RGB) $X_{3,\text{e}}$ already has fallen to $7.55\,10^{-5}$. Our mass loss description leads to the loss of only $0.0002\,M_\odot$ up to then. After the second dredge-up the convective envelope penetrates to depths of almost

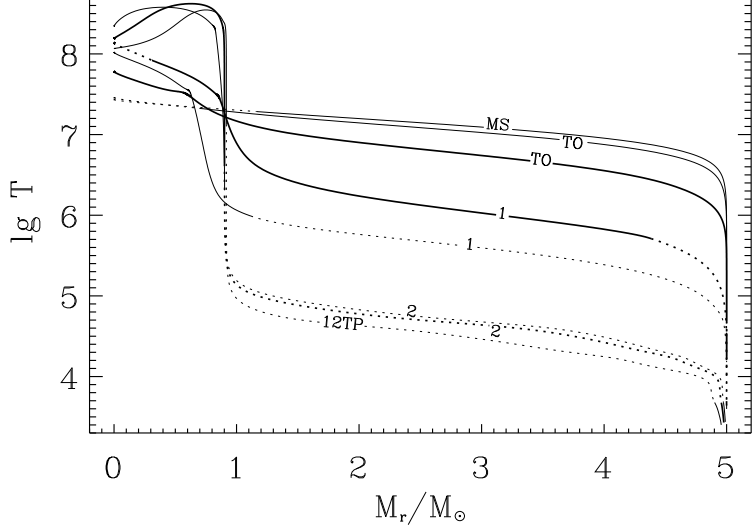

**Fig. 2.** Temperature profile of the 5 $M_\odot$ Pop. I star during its evolution. The models are the same as in Fig. 1 with the addition of two models before and after the second dredge-up ("2")

vanishing $^3$He content and $X_{3,e}$ is dropping to $7.08\,10^{-5}$. Until the end of our calculations it is further reduced by $1\,10^{-7}$ due to $^3$He($^3$He, 2p)$^4$He reactions taking place at the increasingly hotter bottom of the convective envelope.

$g_3$ is loosely defined to be "the fraction of $^3$He that survives stellar processing", without further definition what "survives" means. Here, we define $g_3$ as the ratio of the $^3$He/H abundance between the matter returned to the ISM and that of the initial composition. This implies that the stellar remnant, being inert to further chemical evolution of the galaxy is not included. For the star under discussion $g_3 = X_{3,\mathrm{ef}} X_{\mathrm{ei}} / X_{3,\mathrm{ei}} X_{\mathrm{ef}}$ (final to initial envelope abundance), since effectively all mass is lost at the end of the AGB evolution; the result is $g_3 = 1.20$. Concerning the total amount of $^3$He the ratio final/initial is 0.96. Our results for this star are in good agreement with Galli et al. (1994), although they follow the evolution only up to the end of the first dredge-up. They, too, find that a $5 M_\odot$ star is returning about as much $^3$He as it took from the ISM.

## 2.2. 1.25 $M_\odot$, Pop. II model

As an illustrative example for low-mass stars, we will now discuss the $^3$He evolution of a 1.25 $M_\odot$ Pop. II star, which we followed until the 5th thermal pulse. The calculation was stopped there because we encountered the "Hydrogen Recombination Instability" (Wagenhuber & Weiss 1994b), which we associate with the phase of final envelope ejection. The initial temperature profile (Fig. 3; line labelled "MS") is flat and very similar to that in more massive stars, although temperatures are generally lower. In connection with the much longer main sequence lifetime ($2.7\,10^9$ yrs) this results in a much higher $^3$He abundance outside the regions of complete pp-processing of up to $1.5\,10^{-3}$ (Fig. 4), which, until the end of the main sequence phase, increases further. At the beginning of the first dredge-up, convection is penetrating into the $^3$He-peak, leading to an abundance of $3\,10^{-4}$ in the envelope. In that phase the maximum total $^3$He content ($4.8\,10^{-4}\,M_\odot$) is reached in the star. After the first dredge-up, convection has penetrated deep into the $^3$He-enriched layers and the envelope abundance has grown to $5\,10^{-4}$, which is almost a factor of ten larger than the initial value. This indicates that for low-mass stars the initial $^3$He content is completely irrelevant as long as it is not much larger than solar.

From Fig. 3 it is evident that the effect of core growth is much more important than for more massive stars. Between a model on the subgiant branch (curve "1" with smaller convective envelope) and one at the end of the RGB ("HeF") the hydrogen exhausted ($^3$He-free) core has doubled its size; it will grow by another 15% until the end of the AGB-phase ("5TP"). Due to core growth the total amount of $^3$He compared to the maximum value is halved. The influence of the core helium burning phase is interesting: due to the higher temperatures above the burning core, $^3$He is quickly destroyed between $M_r/M_\odot = 0.48$ and 0.65 but slightly enhanced at $M_r/M_\odot \approx 0.70$ (Fig. 4). Afterwards, when the convective envelope deepens again, this leads to a drop of the envelope $^3$He abundance, which in addition is reduced during the AGB phase. The mechanism is illustrated in Fig. 5: during the TP, the layers between the hydrogen burning shell and the bottom of the convective envelope become much cooler (line 2 in upper panel) than during the quiet interpulse phase (line 1), such that the convective envelope penetrates into regions of much lower $^3$He content (lower panel) and the surface $^3$He abundance is reduced (see inset). After the pulse, a temperature profile is re-established (3) that is similar to that before the pulse, but shifted to higher mass. The corresponding $^3$He profile already displays a beginning reduction of $^3$He from (2) to (3) at $M_r/M_\odot \approx 0.4735$ due to the increasing temperatures. The dilution effect is very small, however, because of the small mass involved compared to the extent of the convective envelope. During the quiescent interpulse phase, $^3$He is slightly increasing at the bottom of the convective envelope. (In more massive stars, where temperatures are higher, it is destroyed at the same location.) The net effect is a change of $\triangle X_3 = -4\,10^{-8}$ per pulse.

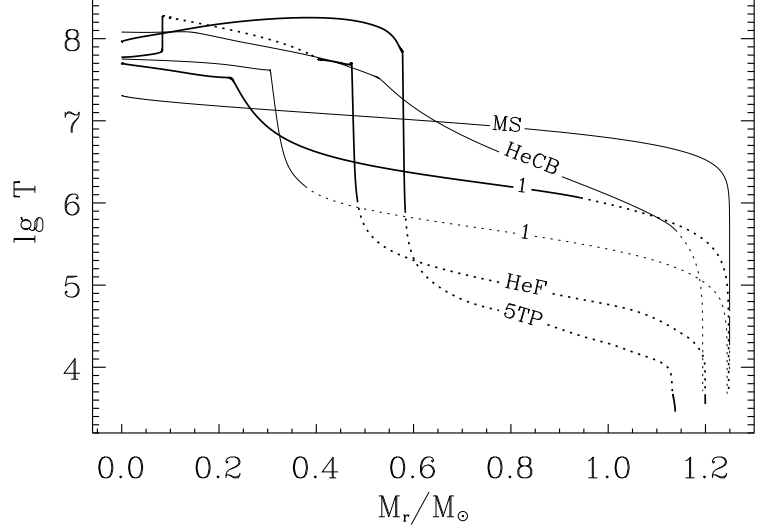

**Fig. 3.** Temperature profile of the 1.25 $M_\odot$ Pop. II star during its evolution until the 5th thermal pulse. Labels along the curves denote models on the main sequence (MS), before and after the first dredge-up (1), during the core helium flash (HeF), central helium burning (HeCB) and after the last thermal pulse (5TP), when the calculations were stopped. Dotted parts mark convective regions

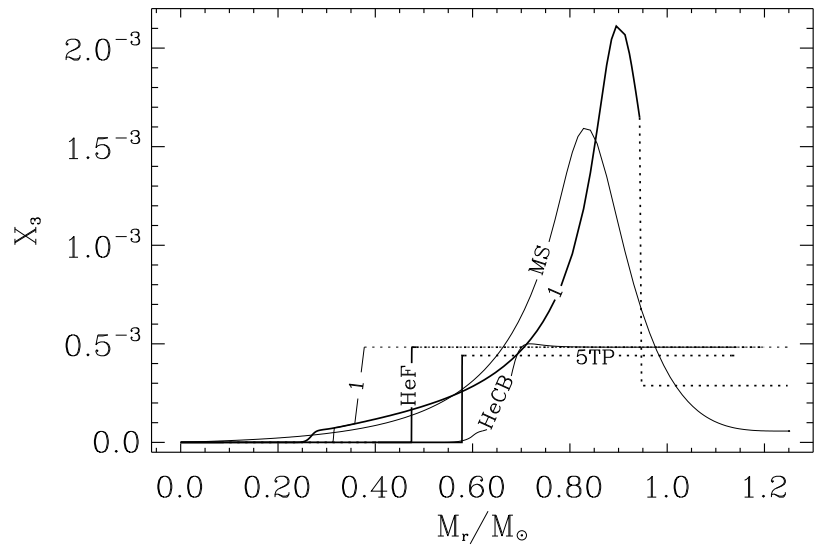

**Fig. 4.** $^3$He profile of a 1.25 $M_\odot$ Pop. II star in various evolutionary phases. Labels have the same meaning as in 3

If the envelope were lost after the first dredge-up, the enrichment of the ISM would be $g_3 = 5.81$; for the case that envelope ejection takes place after the last TP we calculated, it is 3.44. Since 0.15 $M_\odot$ are lost between the end of the first dredge-up and the last thermal pulse, the effective $g_3$ will be $\approx 4$ depending on the mass loss history along the AGB. Of course, in the calculation of these factors the initial $^3$He abundance enters.

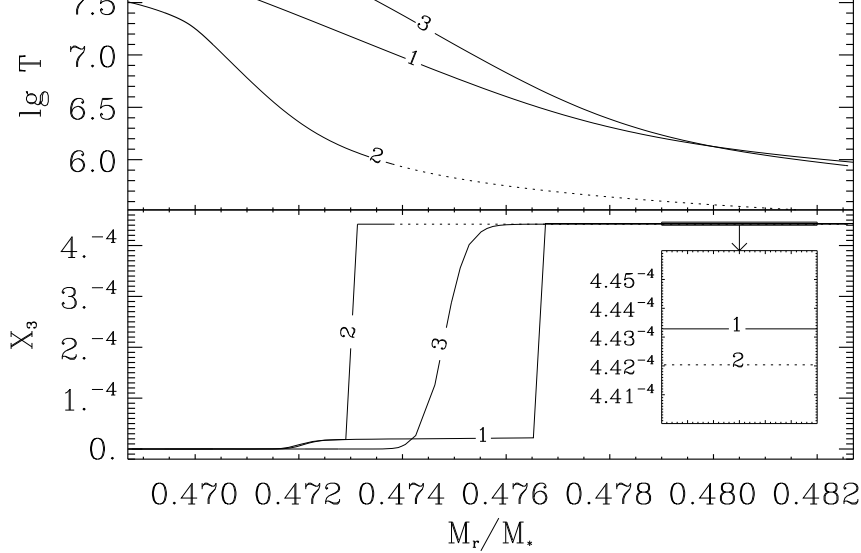

**Fig. 5.** Temperature (upper panel) and $^3$He abundance (lower panel) changes during the first TP of the 1.25 $M_\odot$ Pop. II star. Shown are states before (1), during (2) and after (3) the pulse. The inset demonstrates the net reduction of the surface $^3$He abundance

## 2.3. General results

Tables 1 and 2 summarize our main results for Pop. I and II stars, resp. For each star the first line gives the initial mass and composition and the final $g_3$ value calculated as in the last section under the assumption that practically all envelope mass is lost at the end of the calculation. An exception is the $1 M_\odot$ Pop. I star: for that case mass loss was calculated with $\eta = 0.30$ and was further enhanced to ensure that the complete hydrogen envelope was lost until the late AGB phase. The evolution of this star was followed into the White Dwarf cooling phase. $g_3$ was calculated by averaging the $^3$He abundance of the matter lost.

The two tables list for each star the age (log $t$(yrs)) in the first column. In the second one, we indicate the evolutionary phase the star is approximately in (MS: about half of the main sequence time spent; TO: turn-off; 1DU: first dredge-up; 2DU: second dredge-up; RGB: red giant branch; RGBT: tip of RGB; HeCF: during helium core flash; HeCB: core helium burning; HeSh: Helium shell burning; AGB: early Asymptotic Giant Branch; nTP: n-th thermal pulse; PAGB: post-AGB; WDC: White Dwarf cooling). Then follow the total mass and luminosity in solar units and the effective temperature (log $T_e$(K)). Column 6 gives the envelope mass (in $M_\odot$), which is defined by the layer, where the hydrogen content has fallen to half its photospheric value. The last three columns give the photospheric helium ($Y_e$) and $^3$He ($X_{3,e}$) mass fractions and the average stellar $^3$He abundance ($\langle X_3 \rangle$).

Up to $5 M_\odot$ stars produce more $^3$He during their main sequence phase than they destroy. The effect is particularly strong in the long-lived low-mass stars. After the turn-off, the overall $^3$He-content is decreasing due to the growth of the core and the steepening temperature profile. The surface $^3$He-content always reaches a maximum during the first dredge-up, when the convective envelope penetrates into the former $^3$He core. Especially between 1 and $2 M_\odot$ core helium burning has a major influence on the total $^3$He content, but also on the envelope abundance. As described in the last subsection, the strong temperature increase forces $^3$He to a lower equilibrium abundance. Finally, on the AGB, $X_{3,e}$ decreases slightly (usually less than 1% except for the 7 $M_\odot$ Pop. I and the $1.25 M_\odot$ Pop. II models) due to increasing temperatures at the bottom of the convective envelope. In a test run with $M = 5 M_\odot$ (Pop. I) and a mixing length parameter of 2.0 temperatures at the bottom of the convective envelope were higher by a factor of almost 3. Accordingly, the $^3$He reduction was about twice as effective, but still comparably small. The pulses themselves, we find, influence $^3$He mainly through dilution effects.

Our results for $g_3$ (summarized in Fig. 6) are comparable to those found in the literature. For example, the ratio of initial-to-final $^3$He in our Pop. I 7 and 10 $M_\odot$ stars is 0.70 and 0.54, resp., to be compared with the equivalent value for the 8 $M_\odot$ star of Dearborn et al. (1986) of 0.51. Similarily, our Pop. II 0.8 $M_\odot$ model reaches a maximum envelope abundance similar to that of a model of Charbonnel (1995). Also, the reduction of $X_{3,e}$ after the first dredge-up is very much the same. Next, we can compare the surface mass abundances with Galli et al. (1994) for the 1, 3 and 5 $M_\odot$ models (Pop. I). Again, the agreement is satisfactory, although they appear to have started with a vanishing initial content (see their Fig. 1). Finally, Vassiliadis & Wood (1993) give $^3$He abundances for stars between 0.89 and 5 $M_\odot$ for metallicities typical for the Magellanic Clouds. From their Figs. 22–24 we deduce that the surface $^3$He content after the first dredge-up agrees with our results within a factor of two for most models, except for the most massive ones ($M = 5 M_\odot$), which is obvious because Vassiliadis & Woods used a vanishing initial content and stars

**Table 1.** Results of the calculations for Pop. I models. Ages are given as $\log(t/\text{(yrs)})$; $M_e$ is the mass of the envelope to the point where $X$ is reduced to half the photosperic value; $Y_e$ is the $^4$He envelope mass fraction, $X_{3,e}$ that of $^3$He and $\langle X_3 \rangle$ is the mass fraction averaged over the whole star. ($^3$He fractions are given in units of $10^{-4}$.) All other symbols have their usual meaning; stellar masses and luminosities are in solar units. For the abbreviations of the evolutionary phases (col. 2) see the text

| age | phase | $M(t)$ | $\log L$ | $T_e$ | $M_e$ | $Y_e$ | $X_{3,e} \cdot 10^{-4}$ | $\langle X_3 \rangle \cdot 10^{-4}$ |
|---|---|---|---|---|---|---|---|---|
| $M = 1.0$ | | $X_i = 0.70$ | $Y_i = 0.28$ | $X_{3,i} = 6.02\,10^{-5}$ | | | | $g_3 = 15.58$ |
| 9.657 | MS | 1.000 | 0.000 | 3.761 | 1.000 | 0.2802 | 0.603 | 7.850 |
| 10.062 | TO | 1.000 | 0.387 | 3.685 | 0.842 | 0.2808 | 6.392 | 8.400 |
| 10.085 | 1DU | 0.998 | 1.935 | 3.596 | 0.722 | 0.2935 | 9.322 | 6.728 |
| 10.087 | RGBT | 0.841 | 3.312 | 3.394 | 0.397 | 0.2935 | 9.322 | 4.388 |
| 10.087 | HeCF | 0.779 | 3.400 | 3.366 | 0.313 | 0.2935 | 9.322 | 3.741 |
| 10.087 | HeCB | 0.779 | 1.619 | 3.639 | 0.313 | 0.2935 | 9.322 | 3.741 |
| 10.091 | 1TP | 0.573 | 3.459 | 3.932 | 0.027 | 0.2935 | 9.253 | 0.422 |
| 10.091 | PAGB | 0.552 | 3.304 | 4.139 | 0.004 | 0.2935 | 9.198 | 0.051 |
| 10.091 | PAGB | 0.550 | 3.283 | 4.972 | 0.001 | 0.2935 | 0.000 | 0.003 |
| 10.091 | WDC | 0.550 | 2.604 | 5.000 | 0.000 | 0.2935 | 0.000 | 0.001 |
| 10.091 | WDC | 0.550 | 1.610 | 4.963 | 0.000 | 0.9621 | 0.000 | 0.000 |
| $M = 1.5$ | | $X_i = 0.70$ | $Y_i = 0.28$ | $X_{3,i} = 6.02\,10^{-5}$ | | | | $g_3 = 8.93$ |
| 8.445 | MS | 1.500 | 0.683 | 3.849 | 1.500 | 0.2800 | 0.602 | 1.906 |
| 9.410 | TO | 1.500 | 0.757 | 3.702 | 1.321 | 0.2800 | 0.995 | 4.526 |
| 9.434 | 1DU | 1.500 | 1.085 | 3.664 | 1.299 | 0.2875 | 5.375 | 4.391 |
| 9.459 | 1DU | 1.500 | 2.219 | 3.586 | 1.195 | 0.2889 | 5.321 | 4.231 |
| 9.462 | HeCF | 1.496 | 3.351 | 3.442 | 1.034 | 0.2889 | 5.321 | 3.676 |
| 9.469 | HeCB | 1.496 | 1.725 | 3.641 | 1.009 | 0.2889 | 5.321 | 3.447 |
| 9.479 | AGB | 1.494 | 3.418 | 3.434 | 0.941 | 0.2889 | 5.310 | 3.342 |
| 9.479 | 7TP | 1.488 | 3.580 | 3.400 | 0.926 | 0.2889 | 5.310 | 3.303 |
| $M = 2.0$ | | $X_i = 0.70$ | $Y_i = 0.28$ | $X_{3,i} = 6.02\,10^{-5}$ | | | | $g_3 = 5.02$ |
| 7.081 | MS | 2.000 | 1.193 | 3.957 | 2.000 | 0.2800 | 0.602 | 0.626 |
| 8.951 | TO | 2.000 | 1.417 | 3.915 | 1.778 | 0.2800 | 0.602 | 2.539 |
| 8.995 | 1DU | 2.000 | 1.299 | 3.671 | 1.761 | 0.2829 | 3.114 | 2.469 |
| 9.025 | RGB | 2.000 | 2.154 | 3.609 | 1.691 | 0.2849 | 3.002 | 2.535 |
| 9.033 | HeCB | 1.998 | 2.348 | 3.597 | 1.559 | 0.2849 | 3.002 | 2.333 |
| 9.087 | RGBT | 1.997 | 3.323 | 3.481 | 1.450 | 0.2849 | 3.001 | 2.176 |
| 9.088 | 10TP | 1.991 | 3.619 | 3.429 | 1.422 | 0.2849 | 3.001 | 2.144 |
| 9.088 | 12TP | 1.987 | 3.722 | 3.407 | 1.414 | 0.2849 | 3.001 | 2.135 |
| $M = 3.0$ | | $X_i = 0.70$ | $Y_i = 0.28$ | $X_{3,i} = 6.02\,10^{-5}$ | | | | $g_3 = 2.33$ |
| 5.990 | MS | 3.000 | 1.934 | 4.090 | 3.000 | 0.2800 | 0.602 | 0.442 |
| 8.443 | TO | 3.000 | 2.053 | 3.992 | 2.600 | 0.2800 | 0.602 | 1.241 |
| 8.589 | 1DU | 3.000 | 1.896 | 3.675 | 2.524 | 0.2863 | 1.393 | 1.156 |
| 8.663 | HeCB | 2.999 | 3.309 | 3.520 | 2.442 | 0.2863 | 1.393 | 1.134 |
| 8.663 | 5TP | 2.999 | 3.273 | 3.525 | 2.435 | 0.2863 | 1.393 | 1.131 |
| $M = 5.0$ | | $X_i = 0.70$ | $Y_i = 0.28$ | $X_{3,i} = 6.02\,10^{-5}$ | | | | $g_3 = 1.20$ |
| 7.058 | MS | 5.000 | 2.754 | 4.227 | 5.000 | 0.2800 | 0.602 | 0.372 |
| 7.785 | TO | 5.000 | 2.887 | 4.179 | 4.222 | 0.2800 | 0.602 | 0.517 |
| 7.911 | RGB | 5.000 | 2.974 | 3.951 | 4.211 | 0.2800 | 0.602 | 0.559 |
| 7.913 | 1DU | 5.000 | 2.991 | 3.598 | 4.212 | 0.2807 | 0.741 | 0.577 |
| 7.965 | HeCB | 4.999 | 2.770 | 3.640 | 4.139 | 0.2848 | 0.726 | 0.593 |
| 8.019 | 2DU | 4.971 | 4.282 | 3.418 | 4.066 | 0.2945 | 0.708 | 0.579 |
| 8.019 | 6TP | 4.963 | 4.362 | 3.402 | 4.051 | 0.2945 | 0.707 | 0.578 |
| 8.019 | 11TP | 4.956 | 4.369 | 3.401 | 4.039 | 0.2945 | 0.707 | 0.577 |

**Table 1.** (contd.)

| age | phase | $M(t)/$ $M_\odot$ | $\log L$ | $T_e$ | $M_e/$ $M_\odot$ | $Y_e$ | $X_{3,e} \cdot 10^{-4}$ | $\langle X_3 \rangle \cdot 10^{-4}$ |
|---|---|---|---|---|---|---|---|---|
| | | $X_i = 0.70$ | $Y_i = 0.28$ | $X_{3,i} = 6.02\,10^{-5}$ | | | | |
| $M = 7.0$ | | | | | | | | $g_3 = 0.89$ |
| 7.360 | MS | 7.000 | 3.373 | 4.287 | 7.000 | 0.2800 | 0.602 | 0.354 |
| 7.585 | TO | 7.000 | 3.511 | 3.994 | 5.753 | 0.2800 | 0.602 | 0.392 |
| 7.586 | RGB | 7.000 | 3.130 | 3.617 | 5.753 | 0.2800 | 0.633 | 0.390 |
| 7.586 | 2DU | 7.000 | 3.452 | 3.576 | 5.753 | 0.2802 | 0.566 | 0.417 |
| 7.629 | HeCB | 6.999 | 3.507 | 3.744 | 5.702 | 0.2869 | 0.544 | 0.437 |
| 7.681 | AGB | 6.993 | 4.227 | 3.470 | 5.757 | 0.3013 | 0.512 | 0.421 |
| 7.681 | AGB | 6.983 | 4.473 | 3.426 | 5.924 | 0.3299 | 0.499 | 0.424 |
| $M = 10.0$ | | $X_i = 0.70$ | $Y_i = 0.28$ | $X_{3,i} = 6.02\,10^{-5}$ | | | | $g_3 = 0.72$ |
| 6.395 | MS | 10.000 | 3.779 | 4.398 | 10.000 | 0.2800 | 0.602 | 0.272 |
| 7.243 | TO | 10.000 | 4.003 | 4.331 | 7.868 | 0.2800 | 0.602 | 0.300 |
| 7.288 | 1DU | 10.000 | 3.781 | 3.567 | 7.897 | 0.2835 | 0.446 | 0.340 |
| 7.317 | HeCB | 9.998 | 3.944 | 3.582 | 7.895 | 0.2883 | 0.439 | 0.344 |
| 7.358 | HeSh | 9.987 | 4.459 | 3.474 | 7.657 | 0.2890 | 0.428 | 0.328 |

**Table 2.** Results of the calculations for Pop. II models. Symbols and abbreviations are as in Tab. 1

| age | phase | $M(t)$ | $\log L$ | $T_e$ | $M_e$ | $Y_e$ | $X_{3,e} \cdot 10^{-4}$ | $\langle X_3 \rangle \cdot 10^{-4}$ |
|---|---|---|---|---|---|---|---|---|
| $M = 0.8$ | | $X_i = 0.7499$ | $Y_i = 0.25$ | $X_{3,i} = 5.75\,10^{-5}$ | | | | $g_3 = 12.72$ |
| 1.000 | ZAMS | 0.800 | -0.227 | 3.792 | 0.800 | 0.2500 | 0.575 | 0.575 |
| 9.957 | MS | 0.799 | 0.165 | 3.825 | 0.723 | 0.2500 | 0.575 | 10.870 |
| 10.101 | RGB | 0.798 | 0.854 | 3.744 | 0.599 | 0.2500 | 0.652 | 8.997 |
| 10.112 | 1DU | 0.788 | 2.108 | 3.654 | 0.468 | 0.2552 | 9.969 | 5.713 |
| 10.113 | HeCF | 0.672 | 3.225 | 3.480 | 0.180 | 0.2552 | 9.969 | 2.646 |
| 10.113 | HeCB | 0.672 | 1.640 | 4.003 | 0.180 | 0.2552 | 9.969 | 1.515 |
| 10.116 | 4TP | 0.635 | 2.879 | 3.566 | 0.103 | 0.2570 | 7.327 | 1.163 |
| 10.116 | 5TP | 0.619 | 3.338 | 3.472 | 0.080 | 0.2570 | 7.248 | 0.915 |
| $M = 1.0$ | | $X_i = 0.7499$ | $Y_i = 0.25$ | $X_{3,i} = 5.75\,10^{-5}$ | | | | $g_3 = 9.77$ |
| 1.000 | ZAMS | 1.000 | 0.215 | 3.852 | 1.000 | 0.2500 | 0.575 | 0.575 |
| 9.687 | MS | 0.999 | 0.724 | 3.914 | 0.851 | 0.2500 | 0.575 | 5.473 |
| 9.758 | TO | 0.998 | 1.138 | 3.740 | 0.777 | 0.2500 | 0.765 | 6.041 |
| 9.769 | 1DU | 0.990 | 2.220 | 3.656 | 0.663 | 0.2588 | 6.830 | 4.398 |
| 9.771 | HeCF | 0.918 | 3.193 | 3.516 | 0.433 | 0.2588 | 6.830 | 3.207 |
| 9.771 | HeCB | 0.915 | 1.832 | 3.767 | 0.411 | 0.2588 | 6.830 | 2.140 |
| 9.779 | 2TP | 0.836 | 3.798 | 3.358 | 0.265 | 0.2595 | 5.555 | 1.762 |
| $M = 1.25$ | | $X_i = 0.7499$ | $Y_i = 0.25$ | $X_{3,i} = 5.75\,10^{-5}$ | | | | $g_3 = 7.79$ |
| 1.000 | ZAMS | 1.250 | 0.645 | 3.946 | 1.250 | 0.2500 | 0.575 | 0.575 |
| 9.011 | MS | 1.250 | 0.790 | 3.966 | 1.250 | 0.2584 | 0.575 | 3.497 |
| 9.431 | TO | 1.248 | 1.426 | 3.729 | 1.005 | 0.2502 | 2.879 | 3.734 |
| 9.441 | 1DU | 1.244 | 2.105 | 3.678 | 0.938 | 0.2617 | 4.832 | 3.412 |
| 9.446 | RGBT | 1.200 | 3.126 | 3.553 | 0.727 | 0.2618 | 4.830 | 2.919 |
| 9.456 | HeCB | 1.194 | 2.077 | 3.710 | 0.665 | 0.2618 | 4.830 | 2.150 |
| 9.461 | 5TP | 1.138 | 3.572 | 3.459 | 0.561 | 0.2621 | 4.406 | 2.167 |
| $M = 1.50$ | | $X_i = 0.7499$ | $Y_i = 0.25$ | $X_{3,i} = 5.75\,10^{-5}$ | | | | $g_3 = 5.45$ |
| 1.000 | ZAMS | 1.500 | 0.974 | 4.020 | 1.500 | 0.2500 | 0.575 | 0.575 |
| 8.977 | MS | 1.500 | 1.218 | 4.048 | 1.402 | 0.2500 | 0.575 | 2.319 |
| 9.181 | TO | 1.500 | 1.977 | 3.697 | 1.210 | 0.2593 | 3.453 | 2.447 |
| 9.187 | 1DU | 1.500 | 2.506 | 3.652 | 1.142 | 0.2631 | 3.363 | 2.496 |
| 9.191 | RGBT | 1.489 | 3.003 | 3.591 | 1.037 | 0.2631 | 3.363 | 2.336 |
| 9.191 | HeCB | 1.489 | 2.225 | 3.685 | 1.037 | 0.2631 | 3.363 | 2.278 |
| 9.219 | 1TP | 1.466 | 3.415 | 3.523 | 0.892 | 0.2635 | 3.089 | 1.870 |
| 9.219 | 3TP | 1.456 | 3.544 | 3.495 | 0.872 | 0.2635 | 3.083 | 1.843 |
| 9.219 | 5TP | 1.441 | 3.592 | 3.483 | 0.846 | 0.2635 | 3.077 | 1.805 |

**Table 2. (contd.)**

| age | phase | $M(t)$ | $\log L$ | $T_e$ | $M_e$ | $Y_e$ | $X_{3,e} \cdot 10^{-4}$ | $\langle X_3 \rangle \cdot 10^{-4}$ |
|---|---|---|---|---|---|---|---|---|
| $M = 2.0$ | | $X_i = 0.7499$ | | $Y_i = 0.25$ | | $X_{3,i} = 5.75\,10^{-5}$ | | $g_3 = 3.93$ |
| 1.000 | ZAMS | 2.000 | 1.464 | 4.126 | 2.000 | 0.2500 | 0.575 | 0.575 |
| 8.377 | MS | 2.000 | 1.584 | 4.136 | 2.000 | 0.2954 | 0.575 | 1.144 |
| 8.772 | TO | 1.998 | 1.975 | 3.896 | 1.707 | 0.2500 | 0.575 | 1.775 |
| 8.793 | 1DU | 1.994 | 2.581 | 3.662 | 1.622 | 0.2612 | 2.392 | 1.894 |
| 8.853 | HeCB | 1.988 | 2.316 | 3.721 | 1.423 | 0.2614 | 2.377 | 1.447 |
| 8.866 | AGB | 1.977 | 3.225 | 3.585 | 1.391 | 0.2615 | 2.246 | 1.551 |
| 8.866 | 3TP | 1.960 | 3.662 | 3.505 | 1.351 | 0.2616 | 2.226 | 1.533 |
| 8.866 | 5TP | 1.948 | 3.716 | 3.493 | 1.327 | 0.2617 | 2.223 | 1.514 |
| $M = 3.0$ | | $X_i = 0.7499$ | | $Y_i = 0.25$ | | $X_{3,i} = 5.75\,10^{-5}$ | | $g_3 = 2.49$ |
| 1.000 | ZAMS | 3.000 | 2.106 | 4.250 | 3.000 | 0.2500 | 0.575 | 0.575 |
| 8.186 | MS | 2.999 | 2.309 | 4.227 | 2.585 | 0.2500 | 0.575 | 0.760 |
| 8.307 | MS | 2.998 | 2.536 | 3.984 | 2.549 | 0.2500 | 0.575 | 0.906 |
| 8.388 | TO | 2.994 | 2.824 | 4.002 | 2.263 | 0.2500 | 0.575 | 0.925 |
| 8.405 | 2DU | 2.981 | 3.709 | 3.545 | 2.210 | 0.2539 | 1.426 | 1.052 |
| 8.406 | 3TP | 2.956 | 4.024 | 3.481 | 2.176 | 0.2543 | 1.423 | 1.047 |
| 8.406 | 5TP | 2.947 | 4.088 | 3.466 | 2.160 | 0.2544 | 1.422 | 1.042 |
| $M = 4.0$ | | $X_i = 0.7499$ | | $Y_i = 0.25$ | | $X_{3,i} = 5.75\,10^{-5}$ | | $g_3 = 1.60$ |
| 1.000 | ZAMS | 4.000 | 2.536 | 4.322 | 4.000 | 0.2500 | 0.575 | 0.575 |
| 8.015 | MS | 3.998 | 2.841 | 4.295 | 3.361 | 0.2500 | 0.575 | 0.559 |
| 8.092 | TO | 3.994 | 3.171 | 4.154 | 3.072 | 0.2500 | 0.575 | 0.598 |
| 8.115 | HeCB | 3.992 | 3.155 | 3.790 | 2.978 | 0.2500 | 0.575 | 0.600 |
| 8.116 | 2DU | 3.991 | 3.382 | 3.624 | 2.977 | 0.2512 | 0.976 | 0.641 |
| 8.118 | AGB | 3.949 | 4.177 | 3.485 | 3.034 | 0.2685 | 0.896 | 0.688 |
| 8.118 | 3TP | 3.947 | 4.257 | 3.467 | 3.029 | 0.2694 | 0.896 | 0.687 |
| 8.118 | 5TP | 3.944 | 4.275 | 3.463 | 3.025 | 0.2694 | 0.896 | 0.687 |
| 8.118 | 8TP | 3.939 | 4.303 | 3.457 | 3.016 | 0.2694 | 0.895 | 0.686 |
| 8.118 | 10TP | 3.935 | 4.319 | 3.453 | 3.010 | 0.2695 | 0.895 | 0.685 |
| $M = 5.0$ | | $X_i = 0.7499$ | | $Y_i = 0.25$ | | $X_{3,i} = 5.75\,10^{-5}$ | | $g_3 = 1.23$ |
| 1.000 | ZAMS | 5.000 | 2.859 | 4.373 | 5.000 | 0.2500 | 0.575 | 0.575 |
| 7.712 | MS | 4.998 | 3.056 | 4.342 | 4.148 | 0.2500 | 0.575 | 0.368 |
| 7.834 | TO | 4.997 | 3.284 | 4.211 | 4.126 | 0.2500 | 0.575 | 0.426 |
| 7.898 | HeCB | 4.993 | 3.491 | 4.211 | 3.766 | 0.2500 | 0.575 | 0.432 |
| 7.912 | 2DU | 4.980 | 4.101 | 3.531 | 3.701 | 0.2526 | 0.712 | 0.529 |
| 7.913 | 5TP | 4.958 | 4.310 | 3.483 | 4.003 | 0.2999 | 0.667 | 0.538 |
| 7.913 | 10TP | 4.952 | 4.371 | 3.469 | 3.992 | 0.2999 | 0.666 | 0.537 |
| 7.913 | 18TP | 4.944 | 4.402 | 3.462 | 3.978 | 0.3000 | 0.666 | 0.536 |
| 7.913 | 19TP | 4.944 | 4.624 | 3.414 | 3.977 | 0.3000 | 0.666 | 0.536 |
| $M = 5.0$ | | $X_i = 0.7499$ | | $Y_i = 0.25$ | | $X_{3,i} = 2.9\,10^{-7}$ | | $g_3 = 156.2$ |
| 7.676 | MS | 5.000 | 3.041 | 4.347 | 4.152 | 0.2500 | 0.003 | 0.236 |
| 7.893 | TO | 4.995 | 3.508 | 4.139 | 3.751 | 0.2500 | 0.003 | 0.317 |
| 7.902 | 2DU | 4.990 | 3.821 | 3.580 | 3.726 | 0.2513 | 0.461 | 0.333 |
| 7.903 | AGB | 4.963 | 4.284 | 3.488 | 4.014 | 0.2987 | 0.424 | 0.343 |
| 7.903 | 3TP | 4.961 | 4.303 | 3.484 | 4.010 | 0.2988 | 0.424 | 0.342 |

of that mass approximately preserve the initial abundance. The effect of the second dredge-up is a reduction of $X_{3,e}$ of up to 10%, most expressed for the lowest and highest masses of lowest metallicity. While for the first case this trend agrees with our results, we find no reduction for the latter case. This discrepancy might again be due to the almost vanishing $^3$He content after the first dredge-up in the models of Vassiliadis & Wood. To summarize, in qualitative agreement with the combined results of all previous studies (see also Tab. 3 of Galli et al. 1995 for a compilation of results) we find that stars with $M \lesssim 5 M_\odot$ are net producers of $^3$He, while those more massive are net destroyers. Our $g_3$ values are in general higher than previous ones, partly because of our definition. We also find that $g_3$ is smaller for a lower metallicity (Dearborn et al. 1986).

In addition to the standard calculations, we have added a 5 $M_\odot$ model with an initial $^3$He abundance which was obtained by scaling the solar one with the metallicity. We have chosen this mass value, because for lower masses we already have seen that the final $^3$He abundance is almost independent of the initial one, and because more massive stars are reducing $^3$He anyway.

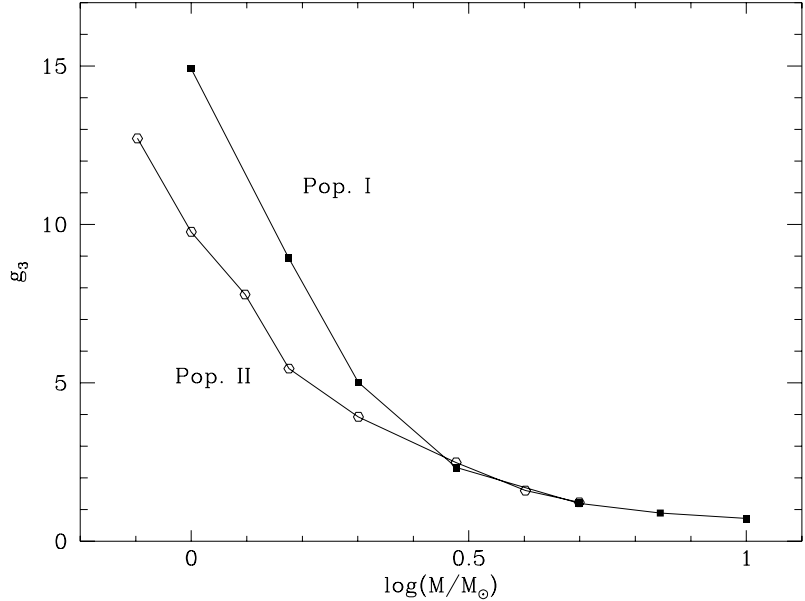

**Fig. 6.** $^3$He survival fraction $g_3$ (as defined in Sect. 3.1) for our models. Solid symbols: Pop. I; open: Pop. II

This model is listed at the end of Tab. 2. The final $X_{3,e}$ is 64% of that for the standard initial $X_3$ content, which is remarkable considering that the initial values differ by a factor of almost 200. Consequently, $g_3$ is 156.2 compared to 1.2. We conclude that for all stellar masses which lead to a net production of $^3$He, the initial abundance is of minor importance for the final envelope abundance.

## 3. $^3$He destruction by diffusion in red giants

Our standard calculations confirm the fact that low-mass stars are net producers of $^3$He. In fact, our $g_3$ values tend to be even higher than previous ones. The problems mentioned in the introduction concerning the chemical evolution of the galaxy or the allowed primordial D abundance therefore persist or are actually more severe. They could be solved if low-mass stars were net destroyers of $^3$He. Hogan (1995) argued that the observed low $^{12}$C/$^{13}$C abundance ratio observed in globular cluster stars is indicative of a significantly reduced $^3$He abundance due to the similar rates of the $^{12}$C(p, $\gamma$)$^{13}$C and $^3$He($^3$He, 2p)$^4$He reactions. His conclusion was that stars with a lower $^{12}$C/$^{13}$C than predicted by standard stellar evolution must also have a lower $^3$He content, and that thus low-mass stars might well be destroying $^3$He. However, contrary to $^{12}$C, $^3$He not only is destroyed, but also is produced - especially at lower temperatures. Therefore, it is not obvious that stars in which $^{12}$C is converted to $^{13}$C will also have $^3$He destroyed and a detailed analysis is worthwhile.

The problem of abundance anomalies in globular cluster red giants has recently been investigated by Denissenkov & Weiss (1996), who showed that additional non-standard mixing processes can very well explain these anomalies, of which the low $^{12}$C/$^{13}$C value correlated with $M_v$ is only one (see this paper for the general problem and more references). Denissenkov & Weiss (1996) described the additional non-standard mixing process by a diffusion equation with the diffusion coefficient $D$ and the mixing depth $\delta m$ as parameters. Although no particular mixing mechanism was assumed, the treatment was motivated by the idea that differential rotation might induce diffusive mixing. It suggests itself to use the same approach and the same parameters that were successful with respect to the $^{12}$C/$^{13}$C – $M_v$ relation and investigate the corresponding $^3$He evolution.

The first model inspected was the 0.8 $M_\odot$ Pop. II star just finishing its first dredge-up and being close to the maximum envelope $^3$He abundance ($\log L/L_\odot = 1.21$). The model was then evolved with the inclusion of diffusion (see Denissenkov & Weiss 1996 for details about the method) and the parameters $D = 5\,10^7\,\text{cm}^{-2}/\text{s}$ and $\delta m = 0.16$, where $\delta m$, the "mixing depth", is a relative mass coordinate being 1 at the bottom of the convective zone and 0 where $X = 0.0$, i.e. at the bottom of the hydrogen shell. $\delta m = 0.16$ corresponds to a depth where 0.27 % of the initial hydrogen content has been burnt. Our diffusive mixing does not penetrate into the molecular weight gradient believed to inhibit rotationally induced mixing. Fig. 7 displays the evolution of $^3$He from the initial model until the end of the calculation ($1.3\,10^8$ yrs later). During that time the luminosity has increased to $\log L/L_\odot = 1.56$. Clearly, the diffusion needed for the carbon isotope anomaly results in a severe destruction of $^3$He. The final abundance is $1.39\,10^{-5}$ compared to $6.99\,10^{-4}$ at the onset of diffusion.

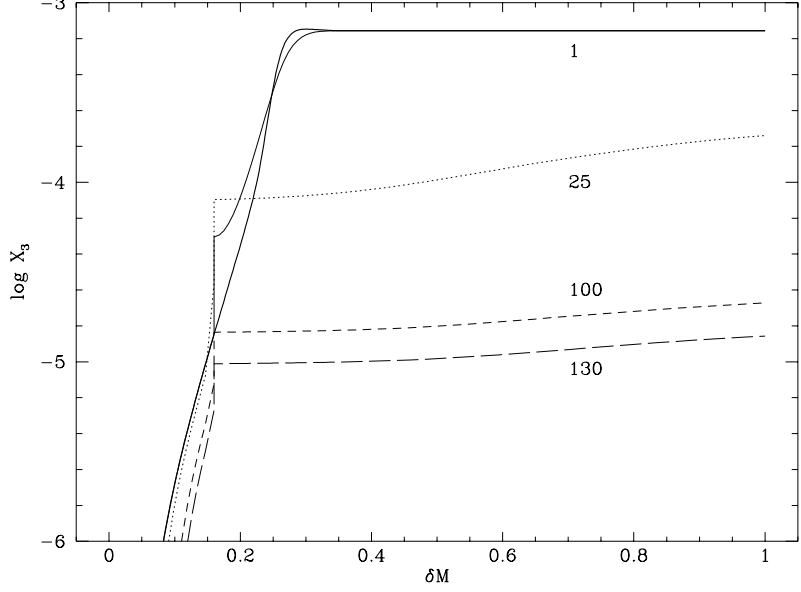

**Fig. 7.** Evolution of $^3$He in the presence of diffusion ($D = 5\,10^7$ cm$^{-2}$/s; $\delta m = 0.16$) for the 0.8 $M_\odot$ Pop. II star. Shown is the $^3$He profile between hydrogen shell and bottom of convective envelope (relative mass coordinate $\delta m$ for four models (solid – dotted – short-dashed – long-dashed lines) at different times, given in $10^6$ yrs. The initial profile is indicated by the thick solid line

**Table 3.** Final $^3$He envelope abundance for different combinations of the parameters $D$ (in cm$^{-2}$/s and $\delta m$ (0.8 $M_\odot$ Pop. II and 1.0 $M_\odot$ Pop. I RGB models). $X_{3,\delta m}$ is the initial $^3$He content of the layer to which diffusion was allowed to penetrate and $\triangle X$ the reduction of hydrogen relative to the envelope abundance (in per cent) at that point; $X_{3,\mathrm{f}}$ denotes the final photospheric abundance of $^3$He

| $D$ | $\delta m$ | $\triangle X$ (%) | $X_{3,\delta m}$ | $X_{3,\mathrm{f}}$ |
|---|---|---|---|---|
| | M=0.8$M_\odot$ | Pop. II | 1$^{\mathrm{st}}$ dredge-up | |
| $5\,10^7$ | 0.16 | 0.3 | $1.47\,10^{-5}$ | $1.39\,10^{-5}$ |
| $5\,10^7$ | 0.15 | 0.4 | $1.03\,10^{-5}$ | $9.63\,10^{-6}$ |
| $5\,10^7$ | 0.13 | 0.7 | $6.85\,10^{-6}$ | $5.42\,10^{-6}$ |
| $1\,10^8$ | 0.10 | 2.6 | $2.12\,10^{-6}$ | $1.36\,10^{-6}$ |
| $5\,10^7$ | 0.18 | 0.1 | $2.49\,10^{-5}$ | $2.61\,10^{-5}$ |
| $1\,10^8$ | 0.20 | .06 | $4.52\,10^{-5}$ | $4.50\,10^{-5}$ |
| | M=1.0$M_\odot$ | Pop. I | 1$^{\mathrm{st}}$ dredge-up | |
| $5\,10^7$ | 0.12 | 0.4 | $1.15\,10^{-5}$ | $1.42\,10^{-5}$ |
| $5\,10^7$ | 0.14 | 0.2 | $1.86\,10^{-5}$ | $2.77\,10^{-5}$ |
| $1\,10^8$ | 0.13 | 0.3 | $1.47\,10^{-5}$ | $1.44\,10^{-5}$ |

In Tab. 3 we have collected the results of several different parameter pairs (first four rows). We find that (i) the final $^3$He abundance is almost equal to the initial one in that layer down to which diffusion was allowed to penetrate; (ii) that diffusion is always fast enough to ensure complete mixing of the envelope; and (iii) that the penetration depth is deep enough such that $^3$He can reach the corresponding equilibrium abundance within the mixing time. Similar calculations were performed for the 1.0 $M_\odot$ Pop. I model with very similar results (Tab. 3; last group of rows).

Next we investigated whether it would be possible to get $^3$He destruction without changing the $^{12}$C/$^{13}$C ratio. However, we found that this is not possible (cf. Tab. 3; last two rows for the 0.8 $M_\odot$ star) as long as one makes the reasonable assumption that carbon and helium diffusion have the same timescale. The reason is that a siginificantly lower $^3$He abundance (compared to the envelope value) is reached only in layers that also have a non-solar $^{12}$C/$^{13}$C ratio (Fig. 8). Only a modest reduction of $^3$He at constant $^{12}$C/$^{13}$C is possible with a very fine-tuned diffusion depth of $\approx 0.25$. But such a choice would result only in a correction to the $^3$He-overproduction, but not in a net destruction. The close correlation of $^3$He with the $^{12}$C/$^{13}$C ratio is further demonstrated in Fig. 9, where the evolution of both quantities is shown.

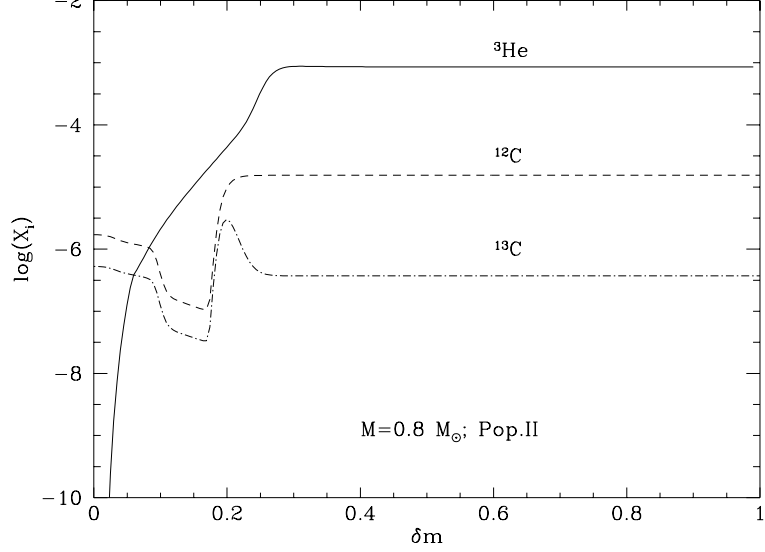

**Fig. 8.** Initial run of $^3$He, $^{12}$C and $^{13}$C in the diffusive region of the 0.8 $M_\odot$ model

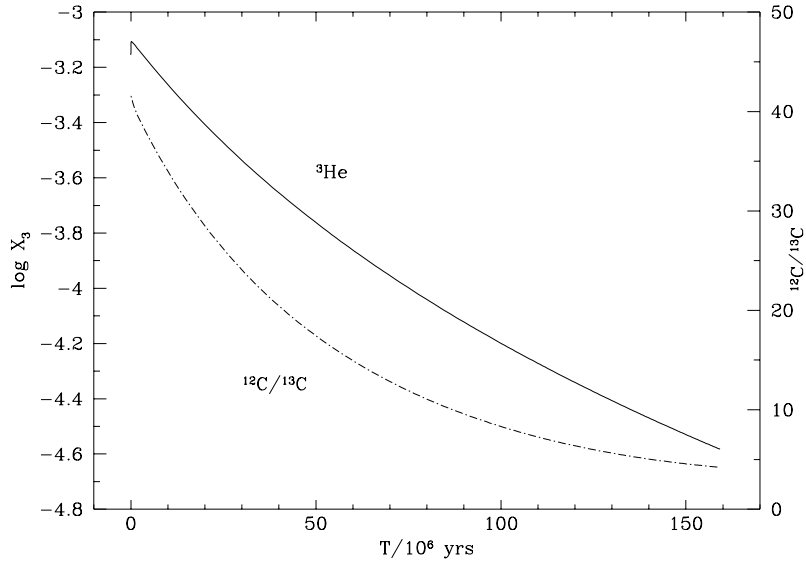

**Fig. 9.** Evolution of envelope $X_3$ and $^{12}$C/$^{13}$C ratio with time in the 0.8 $M_\odot$ star for $D = 1\,10^8$ cm$^{-2}$/s and $\delta m = 0.20$ (very shallow mixing)

Charbonnel (1995) recently and independently of us has published results of a similar investigation for 0.8 and 1.0 $M_\odot$ stars of low metallicity (followed up to the RGB tip). Her quantitative results and final conclusion agrees with ours; she, too, finds that $^3$He and $^{12}$C/$^{13}$C are strongly correlated and both are reduced by additional mixing in RGB stars. She finds a net $^3$He reduction for the $Z = 10^{-3}$ cases, but not so for the $Z = 10^{-4}$, $M = 0.8 M_\odot$ model. A more qualitative "conveyor-belt model" was used by Wasserburg, Boothroyd & Sackmann (1995) to support Hogan's (1995) conjecture. They deduce a reduction of $^3$He by a factor of 2. We therefore conclude this section by summarizing that a mixing mechanism beyond ordinary convection in Red Giant envelopes, which leads to the well-observed carbon isotope anomalies is extremely likely to revert the envelope $^3$He content to the initial value or even below. Our parametric calculations indicate final $g_3$ values of 0.2 or less.

In this paper we have investigated the evolution of stars of low and intermediate mass ($1.0\cdots 10.0 M_\odot$ for Pop. I and $0.8\cdots 7.0 M_\odot$ for Pop. II) with respect to their $^3$He production. The computations included the AGB phase with thermal pulses. Under standard assumptions we found in agreement with earlier work:

1. Stars are net producers of $^3$He for $M \lesssim 5 M_\odot$; $g_3$-factors are summarized in Fig. 6;
2. $g_3$ increases with decreasing mass and increasing metallicity;
3. $\langle g_3 \rangle$ (mean $g_3$ over an IMF with slope $-2.7$) is 11.6 for Pop. I and 7.5 for Pop. II;
4. $^3$He is produced in great quantities during the main sequence phase and mixed to the surface during the first dredge-up;
5. subsequent evolutionary phases lead to the reduction of $^3$He abundances due to core growth, dilution phases and hot bottom processing; however, this reduction is only a small effect compared to the initial enhancement in low-mass stars;
6. the absolute final $^3$He abundances are almost independent of the initial abundance; this is particularily true for the lower masses;
7. although there is a good qualitative agreement between the various published papers, including the present one, quantitatively the agreement is not better than a factor of two.

To investigate a suggestion by Hogan (1995), we also carried out several computations that included a parameterized description of diffusion between the bottom of the convective envelope and the hydrogen shell. These calculations show that

1. for parameters needed to explain observed $^{12}$C/$^{13}$C anomalies in Red Giants, additional diffusion after the first dredge-up leads to a net destruction of $^3$He even in low-mass stars;
2. if we assume that for stellar masses below 2 or 3 $M_\odot$ $g_3$ is as low as 0.1 (as indicated by our diffusion calculations), the average $g_3$ can be reduced to 0.98 or 0.31, resp.;
3. a $^3$He reduction in the envelope cannot be achieved without simultaneous $^{12}$C/$^{13}$C values close to CN-equilibrium values.

We therefore have confirmed Hogan's suggestion and an independent work by Charbonnel (1995), which leads us to recognize that low-mass stars contrary to standard evolution can well be net destroyers of $^3$He. To quantify the effect for the chemical evolution of the galaxy one first has to have solid statistics about carbon (and similar) isotope/element anomalies in low-mass Red Giants. Clearly, not all stars will destroy $^3$He, as is evident from the observation of high $^3$He in the Planetary Nebula NGC 3242 (Rood, Bania & Wilson 1992)[1]. We follow Charbonnel (1995), who stated that this can be explained by a progenitor mass of more than 2 $M_\odot$ for NGC 3242. For such masses, extra mixing is not needed to explain observed $^{12}$C/$^{13}$C abundance ratios. If this is a general property, stars below 2 $M_\odot$ would be net destroyers of $^3$He, between 2 and 5 $M_\odot$ they would produce $^3$He by a factor of 2–3, and beyond 5 $M_\odot$ they again would reduce $^3$He in the ISM. The integrated $g_3$ would then be slightly lower than 1, which could explain the similarity of primordial, pre-solar and ISM $^3$He abundances. However, it would not change the argument against a high primordial D abundance, because one would need $\langle g_3 \rangle \lesssim 0.1$ to explain pre-solar abundances (except if one assumes that the pre-solar ISM was heavily polluted with massive star debris). We leave it to the modellers of galacto-chemical evolution to study the effect of our results.

*Acknowledgements.* We are grateful to J. Truran for initiating this study and for additional helpful comments and discussions. P.A.D. acknowledges financial support through the RBRF grant No. 95-02-05014-a. Part of the work of A.W. and J.W. was supported by the "Sonderforschungsbereich 375 für Astro-Teilchenphysik" der Deutschen Forschungsgemeinschaft.

---

[1] Such observations are also contradicting a nuclear explanation for the galactic evolution of $^3$He (Galli et al. 1994).